\documentstyle[multicol,prd,aps,epsf,epsfig]{revtex}

\def \a{\alpha}
\def \b{\beta}
\def \l{\lambda}

\def \be{\begin{equation}}
\def \ee{\end{equation}}
\def \o{\omega}
\def \O{\Omega}

\begin{document}

\title{From brane assisted inflation to quintessence through a single scalar
field}

\author{A. S. Majumdar\footnote{Electronic address: archan@boson.bose.res.in}}

\address{S. N. Bose National Centre for Basic Sciences,
Block JD, Sector III, Salt Lake, Calcutta 700098, India}

\maketitle

\begin{abstract}

We explore within the context of brane cosmology whether it is possible
to obtain both early inflation and  accelerated
expansion during the present epoch through the dynamics 
of the same scalar field in an exponential potential. Considerations from
successful inflation and viable radiation and matter dominated eras impose
constraints on the parameters of the potential. We find that the 
additional requirement of 
late time quintessence behaviour in conformity with present observations
necessitates the inclusion of two exponential terms in the potential.

PACS number(s): 98.80.Cq

\end{abstract}

\begin{multicols}{2}

\section{Introduction}

The idea that our universe is a brane embedded in higher dimensional space
has received considerable attention in recent years\cite{rubakov}. This has 
been motivated by solutions of string theory where all matter and gauge fields
are confined to the 3-brane, whereas gravity can propagate in the bulk. In
these schemes the extra dimensions need not be small or compact, a radical
departure from the standard Kaluza-Klein scenario, and the fundamental Planck
scale could be significantly smaller than our effective four dimensional
Planck scale $m_{pl}$\cite{randall}. A lot of effort is currently being devoted to
understand the cosmology of such a brane world scenario\cite{cline}.

The most important feature that distinguishes brane cosmology from the standard
scenario is the fact that at high energies the Friedmann equation is modified
by an extra term quadratic in energy density $\rho$\cite{cline}. It is expected that
the implications of such a modification would be profound for the inflationary
paradigm. Recent measurements of the power spectrum of CMB anisotropy\cite{bernandis}
provide a strong justification for inflation\cite{lidsey}. It has been 
realized that the brane world scenario is more suitable for inflation with
steep potentials because the quadratic term in $\rho$
increases friction in the inflaton field equation\cite{maartens}. This feature has been 
exploited to construct inflationary models using both large inverse power 
law\cite{huey} as well as steep exponential\cite{copeland} potentials for
the scalar field.

A common ingredient in these models is that reheating is supposed to take
place through gravitational particle production. The conventional reheating
mechanism through decay of the inflaton cannot be implemented in these models
because the saclar field potential does not have a minimum near the scale
of inflation. At the end of inflation, the scalar field equation becomes
kinetic energy dominated because of the steepness of the potential. This
energy is redshifted rapidly and radiation domination ensues. The detailed
aspects of the scenario of gravitational particle production has been 
debated in Refs.\cite{ford}. The condition that radiation domination sets
in before nucleosynthesis is used to impose constraints on the parameters
of brane inflation models\cite{huey,copeland}. A definite prediction of these
models is the parameter independence of the spectral index of scalar
density perturbations\cite{maartens,huey,copeland}.

The continued rolling down the potential slope of the inflaton field raises
interesting questions about its late time behaviour. Preliminary analyses
support the idea that the same scalar field can provide inflation at early
times, and behave as a quintessence field at late times\cite{huey,copeland,peebles}.
Recent observations of distant supernovae and galaxy clusters seem to suggest
that our universe is presently undergoing a phase of accelerated expansion\cite{perlmutter},
indicating the dominance of dark energy with negative pressure in our present
universe. The idea that a slowly rolling scalar field provides the dominant
contribution to the present energy density has gained prominence in recent
times\cite{ratra}. The crucial feature for viability of this theme is the
existence of late time attractor solutions for a wide range of initial
conditions for certain types of potentials\cite{zlatev}. The possibility of
obtaining quintessence through tracker-type potentials has been analysed
through a variety of power law or exponential potentials, and also combinations
of these\cite{copeland2}, though a recent study\cite{doran} seems to 
disfavour models with large inverse power law potentials.

In this paper we consider a potential consisting of a general combination of two
exponential terms given by
\begin{equation}
V(\phi) = V_0[A{\rm exp}(-\alpha\phi/m_{pl}) + B{\rm exp}(-\beta\phi/m_{pl})]
\end{equation}
Such a potential has been recently claimed\cite{barreiro} to conform to all
the observational constraints till date on quintessence for certain values
of parameters. Our goal is to check in detail whether a viable scenario of
brane world inflation, and a feasible late time quintessence solution both
work out with this potential. We impose all the requisite constraints that
come into play in inflation, during the intermediate radiation and matter
dominated eras, and finally from the present epoch of accelerated expansion.
Balancing the dynamics between such disparate scales requires, as expected,
a tuning of one parameter in the potential. This is just a rephrasing of the
cosmological constant problem which cannot be addressed through this mechanism.
Nonetheless, our results indicate that such a unified scheme is possible with 
certain restrictions on the allowed values for the potential parameters. Our
analysis serves to highlight the inevitability of requiring two exponential 
terms and also to rule out certain specific categories (formed by the 
selection of particular combinations of values and signatures of parameters) 
of the above potential. 

\section{Inflation in the brane world scenario}

To begin with, let us note certain points regarding the parameters in potential
(1). First, one of the constants $A$ or $B$ can be absorbed in $V_0$. Secondly,
$V(\phi)$ is invariant under the transformation $\a \to -\a, \b \to -\b, 
\phi \to -\phi$. So without loss of generality we can set $B=1$, $\a < 0$, and
$\phi_i > 0$, where $\phi_i$ is some initial value of $\phi$. Further, we 
assume that either of the two conditions, (a) $-\a >> |\b|$, or (b) $|A| >> 1$,
for $\a/\b \approx +1$ hold. This makes the first term in $V(\phi)$ dominate 
except for very small or negative values of $\phi$, thereby enabling us to 
make use of the analysis of Copeland {\it et al.}\cite{copeland} during the
early inflationary era.

The brane world dynamics is governed by the modified Friedmann equation
\begin{equation}
H^2 = {8\pi \over 3m_{pl}^2}\rho\biggl(1 + {\rho \over 2\l}\biggr)
\end{equation}
where, for reasons of simplicity, we have set the contributions from bulk
gravitons and higher dimensional cosmological constant to 
zero\cite{maartens,huey,copeland}. The brane tension $\l$ defines the scale
below which the quadratic correction begins to lose importance and standard
Friedmann evolution is recovered. The scalar field obeys the equation
\be
\ddot{\phi} + 3H\dot{\phi} + V'(\phi) = 0
\ee
As long as the condition $\rho/2\l >1$ holds, the quadratic term in Eq.(2)
dominates and contributes to increased friction in the scalar field equation 
(3). Thus during inflation, one can define a modified slow role parameter
\be
\epsilon(\phi) \equiv {m_{pl}^2\over 8\pi}\Biggl({V'(\phi)\over V(\phi)}
\Biggr)^2 {2\l \over V(\phi) + 2\l}
\ee 

If the brane energy momentum tensor is dominated by the energy density of
the scalar field, inflation ensues when $\epsilon <1$. The condition for
brane assisted inflation\footnote{The term ``assisted inflation'' was coined 
in the context of standard Friedmann cosmology where cumulative effects of
multiple scalar fields in exponential potentials give rise to 
inflation\cite{liddle}. However, in this paper we use the term ``brane 
assisted inflation''  simply to signify the effect of the quadratic energy
density correction towards helping slow roll.} is thus weakened compared 
to the standard case, as here
$p < -{2\over 3}\rho$ guarantees acclerated expansion, i.e., $\ddot{a} > 0$.
In the slow roll expansion, the number of e-foldings $N$ is given by
\be
N \simeq -{8\pi \over m_{pl}^2}\int_{\phi_i}^{\phi_e} {V(\phi)\over V'(\phi)}
\Biggl({2\l + V(\phi)\over 2\l}\Biggr)d\phi
\ee
where $\phi_e$ denotes the value of $\phi$ at the end of inflation, 
($\epsilon \approx 1$). Assuming $V >> \l$ during inflation, one obtains
\be
V_0 \simeq {\l \a^2 \over 4\pi A}{\rm exp}\biggl({\a \phi_e \over m_{pl}}\biggr)
\ee
and
\be
V_e \simeq {\l \a^2 \over 4\pi}
\ee
Hence, the number of e-foldings $N$ can be written as
\be
N \simeq {\rm exp}\Biggl({-\a (\phi_i - \phi_e) \over m_{pl}}\Biggr) - 1 \simeq
{4\pi \over \l \a^2}(V_N - V_e)
\ee
where the value of the potential at $N$ e-foldings from the end of inflation
is given by
\be
V_N = V_e(N+1)
\ee
Following Refs.\cite{maartens,copeland} one can define the amplitude of 
density perturbations as
\be
A_s^2 \simeq {64\pi V^4(\phi) \over 75 \a^2 \l^3}
\ee
Using the COBE normalization $A_s = 2\times 10^{-5}$\cite{bunn}, one gets
\be
\l \simeq \Biggl({10^{15}{\rm GeV} \over (-\a/\sqrt{8\pi})^{3/2}}\Biggr)^4
\ee
Though we have assumed $V(\phi) >> \l$ during inflation, consistency demands
that the scalar field be confined to the brane at all times. One has to ensure
that $V_{max} < (M_5)^4$, where $M_5$ is the five dimensional Planck
scale, related to $m_{pl}$ and brane tension $\l$ by
\be
m_{pl} = \sqrt{{3\over 4\pi}}\Biggl({(M_5)^2 \over \sqrt{\l}}\Biggr)M_5
\ee
The above consistency condition ($V_{max} < (M_5)^4$) leads to the constraint
\be
{-\a(\phi_{max}-\phi_e)\over m_{pl}} \le 13.8
\ee
Using Eqs.(8) and (13) one can see that the maximum number of e-foldings 
possible is given by
\be
N_{max} \simeq {\rm exp}\Biggl({-\a (\phi_{max}-\phi_e) \over m_{pl}}\Biggr)
\approx 10^5
\ee

We recover the results of brane assisted inflation presented in 
Ref.\cite{copeland} since our model reduces to theirs during inflation. It 
should be mentioned here that two generic predictions of brane inflation, (the
spectral index $n_S \approx 0.92$, and the ratio of the amplitude of tensor
to scalar perturbations, $A^2_T/A^2_S \approx 0.03$),  are independent of the
slope $\a$ of the exponential potential\cite{maartens,copeland,langlois}.
However, to produce an observationally acceptable value for $n_S$ and
$A^2_T/A^2_S$, one needs a very high power of $\phi$ for inverse power law 
potentials\cite{huey}.

\section{Constraints from radiation and matter dominated eras}

The scalar field potential at the end of inflation is given by $V_e$ (Eq.7).
At this stage the brane correction to Friedmann equation is dominant for
$\a^2 > 8\pi$. The field $\phi$ continues to roll down in the absence of
any minimum for the potential at this scale. Thus reheating can take place
only through gravitational particle production\cite{ford}. Assuming this to 
be the case, the energy density in radiation at the end of inflation is 
given by
\be
(\rho_R)_e \simeq g\Biggl({10^{11}{\rm GeV} \over -\a/\sqrt{8\pi}}\Biggr)^4
\ee
where $g$ is the number of fields which produce particles at this stage. The
ratio of radiation density to scalar field density is 
$(\rho_{R})_e/(\rho_{\phi})_e\approx g(10^{-17})$\cite{copeland}. 
Radiation red shifts as $\rho_R \propto a^{-4}$
and competes with the scalar field energy density  $\rho_{\phi}$ for domination.
The equation state for $\phi$ after inflation is $\o_{\phi} \ge -2/3$. 
$\rho_{\phi}$ falls off starting from $\rho_{\phi} \propto a^{-1}$ (when brane
effects are most important) to $\rho_{\phi} \propto a^{-6}$ (complete kinetic
domination). In the rather unlikely case of the former evolution throughout
the time until $\rho < 2\l$, the onset of radiation domination gets delayed.
Radiation domination ensues after the scale factor has expanded by an extra
factor of $(V_e/2\l)^{5/3} \simeq (-\a/\sqrt{8\pi})^{10/3}$\cite{copeland}
over and above the factor of around $10^7$ by which it would have expanded
had kinetic domination of the scalar field set in immediately at the end of
inflation. The temperature of radiation at the onset of radiation domination
is given in this scenario by
\be
T_{RD} \simeq {10^4 {\rm GeV} \over (-\a /\sqrt{8\pi})^{13/3}}
\ee
Radiation density in the universe must dominate before nucleosynthesis, i.e.,
$T_{RD} \ge 1{\rm MeV}$. Hence, one obtains an upper bound on $\a$, i.e.,
$\a \le 10^2$. In practice, however, this scenario is extremely unlikely
because the steep nature of the potential will quickly force $\rho_{\phi}$
to be dominated by kinetic energy ($\o_{\phi} \approx 1, \> \dot{\phi}=a^{-3}$),
and the above bound should be interpreted in a loose sense. Keeping this 
caveat in mind, one could still calculate the number of e-foldings encountered
by the $\phi$ field from the end of inflation to nucleosynthesis, which is
given by
\be
{\rm exp}\Biggl({-\a (\phi_e -\phi_{nuc}) \over m_{pl}}\Biggr)|_{K.D.} \simeq
{10^{21}\a^{10} \over (8\pi)^5}
\ee
In the more likely case of brane effects continuing to play a role in the 
dynamics for some time after the end of inflation, a more plausible solution
for $\phi$ is of the ``tracker type''\cite{zlatev}  where $\o_{\phi} \simeq \o_r = 1/3$. In
this case it turns out that
\be
{\rm exp}\Biggl({-\a (\phi_e -\phi_{nuc}) \over m_{pl}}\Biggr)|_{tracker} 
\simeq \Biggl({a_{nuc} \over a_e}\Biggr)^2
\ee

Before proceeding further, one needs to ensure that the temperature of
radiation density at the end of inflation given by Eq.(16) does not exceed
$10^9{\rm GeV}$, the temperature at which thermal production of gravitinos
might be significant\cite{sarkar}. It can be checked from Eq.(16) that
$T_e \le 10^9 {\rm GeV}$ implies $-\a \ge 10^2$. Using this bound together
with the requirement for the onset of radiation domination before 
nucleosynthesis, we set the value of $\a$ to be
\be
-\a \approx O(10^2)
\ee
Recently, it has been claimed that the requirement that reheat temperature 
after inflation be bounded by production of excess gravitons limits the
brane tension, i.e., $\lambda \ge 10^{11}{\rm Gev}^4$\cite{allahverdi}. Our
choice of $\a$ is consistent with this bound.

Exponential potentials are known to yield tracker solutions. Here we look for
a late time attractor or scaling solution where the scalar field dynamics
mimics that of the dominant background fluid (radiation or matter) with an 
approximately constant ratio between their energy densities.
The tracking condition\cite{zlatev} requires that
\be
{V'(\phi)\over V(\phi)} \simeq {m_{pl}^{-1}\over \sqrt{\O_{\phi}}} 
\ee
holds at various stages of evolution. Eq.(20) can be used to check the 
consistency of the constraint (19) at different stages, for example, at
$t_{nuc}$ when the recently obtained bound 
$(\O_{\phi})_{nuc} < 0.045$\cite{bean} needs to be satisfied. The existence
of tracking behaviour can be determined by the quantity\cite{zlatev}
\be
\Gamma \equiv {V''(\phi)V(\phi)\over (V'(\phi))^2}
\ee
If $\Gamma$ stays nearly constant, then a solution converges to a tracking one.
For our model it is easy to verify that $\Gamma =1$ whenever either of the two
terms dominate in Eq.(1). (With our choice of conditions (a) or (b), we want
the first term to dominate during inflation, as well as throughout the
radiation and matter dominated eras. The transient regime when both the terms
play equitable roles in the dynamics will be discussed in the next section). 

The exact evolution of the $\phi$ field will depend on its energy fraction
$\O_{\phi}$ and its equation of state parameter $\o_{\phi}$ which varies with
time all the way from $\o_{\phi} \approx 1$ (kinetic domination) 
at some stage after
inflation to $\o_{\phi} \to -1$ (quintessence) during the present epoch.
Analysis of recent CMB data constrains $\Omega_\phi \le 0.39$ 
 during the radiation dominated epoch\cite{bean}.
Numerical analysis\cite{johri} suggests that from $t_{nuc}$ to $t_{eq}$
$\O_{\phi}$ rises slowly staying nearly constant around $0.2$. Assuming
$\o_{\phi}$ tracks the behaviour of radiation (i.e., $\o_{\phi} \sim 1/3$)
up to the era of matter-radiation equality, one obtains
\be
{\rm exp}\Biggl({-\a(\phi_{nuc}-\phi_{eq})\over m_{pl}}\Biggr) \simeq 
\Biggl({a_{eq}\over a_{nuc}}\Biggr)^2
\ee
From $t_{eq}$ to the present era the field $\phi$ experiences a few more
e-foldings (${\rm exp}[-\a(\phi_{eq}-\phi_{now})/m_{pl}] \approx O(1)$). During
galaxy formation $\O_{\phi} \le 0.5$, and hence, the dynamics should be such
that $-0.5 < \o_{\phi} < -1/3$ during this era\cite{johri}. These are 
approximate results in the sense that for more accurate results one has to
solve the equations of motion for $\phi$ and $H$. Nevertheless, our use of
average values for $\o_{\phi}$ during different eras is justified as our
purpose here is to estimate the available parameter space for $\phi$ before
$\O_{\phi}$ domination sets in around the present epoch. Or, in other words
we estimate the total magnitude of rolling down the effective slope $\a$ of
the potential experienced by $\phi$ until the present phase of accelerated
expansion influenced by the second term in Eq.(1) is arrived at.

\section{Accelerated expansion during the present epoch}

With a high level of confidence present observations suggest that our universe
has entered an era of accelerated expansion driven by a cosmological constant
or energy density of a scalar field with $\O_{\phi} \approx 0.7$\cite{perlmutter,bahcall}.
Our model with effective slope $\a$ describing the dynamics from inflation to
matter domination is unable to account for a second period of accelerated
expansion. This is because the late time attractor solution used here has
$\o_{\phi} \approx \o_m$. Although recently it has been claimed that a viable
model of quintessence could work with a potential having a single exponential
term\cite{cline2}, the allowed value for the slope of the potential is two
orders of magnitude smaller than  the value of $\a$ that needs to be used
by us. It is known that exponential potentials also allow
for another kind of late time attractor solution, viz., $(V'/V)^2 < 3(\o +1)$
and $\o_{\phi} \approx -1 + (V'/V)^2/3$\cite{zlatev,barreiro}. To achieve such
a solution, we invoke the second term in our potential (1), which plays a 
dominant role in the dynamics for small values of $\phi/m_{pl}$. The potential
$V(\phi)$ has a minimum if $A\a/\b < 0$. The dynamics in this case is
quite different from the case when $\phi$ rolls down monotonically. We will
analyze the two cases separately. 

Let us first consider the case when $A\a/\b > 0$, and the potential at present
is dominated by the second term, i.e.,
\be
V_{now} \simeq {\l\a^2\over 4\pi A}{\rm exp}\Biggl({\a\phi_e\over m_{pl}}\Biggr){\rm exp}\Biggl({-\b\phi_{now}\over m_{pl}}\Biggr)
\ee
It can be seen from Eq.(21) that the tracking condition ($\Gamma =1$) is 
satisfied. However, in the transient regime when both the terms in Eq.(1) are
of comparable magnitude, one obtains $\Gamma = 1$ only for $\a/\b \approx O(1)$.
Since early universe dynamics constrains $-a \approx O(10^2)$, acceptable
values for $\b$ in this case would violate the late time attractor requirement
of $\o_{\phi} \approx -1 + \b^2/3$. So for this scheme to work, the present
universe should be well out of the transient regime. In other words, the 
validity of Eq.(23) should be accurate. 

The value of the potential at present $V_{now}$ should be equal to the present
energy density, $\rho_c \approx 10^{-47}({\rm GeV})^4$. Combining Eq.(17) or
(18) with Eq.(22)
 one could set the average value of the quantity $[-\a(\phi_e -
\phi_{now})/m_{pl}] \approx O(60)$. If we set the value of $\phi_{now}/m_{pl}
\approx O(1)$, then equating $V_{now}$ with $\rho_c$ and using Eq.(11) one
obtains
\be
A \simeq {10^{80}{\rm exp}(\a-\b)\over \a^4}
\ee
That we require such a large value for $A$ does not come as a surprise, since
we have reached the present energy density starting from a scale
$V_e \approx O(10^{15}/\a){\rm GeV}$.

We now analyse the case if $A\a/\b < 0$. Here one gets a minimum for the
potential with
\be
{\phi_{min}\over m_{pl}} = {{\rm ln}(-A\a/\b)\over \a -\b}
\ee
and
\begin{eqnarray}
V_{min} &=& {\l\a^2\over 4\pi A}{\rm exp}\Biggl({\a\phi_e\over m_{pl}}\Biggr)
\Biggl(A(-A\a/\b)^{-\a/(\a-\b)} \nonumber \\
&+& (-A\a/\b)^{-\b/(\a-\b)}\Biggr)
\end{eqnarray}
For this case one does not require the second attractor solution, and hence
no corresponding restriction on the value of $\b$. Numerical integration in
Ref.\cite{barreiro} confirms that when $\phi$ reaches the minimum of the 
potential, the effective cosmological constant $V_{min}$ takes over and
oscillations are damped, thus driving the equation of state towards
$\o_{\phi} = -1$. For our model, the requirement that $V_{min} \approx V_{now}$
can be satisfied for $\phi_{min}/m_{pl} \approx O(1) \approx A$, and 
$-\a \approx 10^2 \approx \b$. However, again setting $[-\a(\phi_e -
\phi_{now})/m_{pl}] \approx O(60)$, the tuning required in this case is 
given by
\be
{-\a \over \b} \simeq 1 - O(10^{-60})\a^3
\ee

We thus find that the basic requirements for viable scenarios of quintessence
is possible for certain choices of the parameters. It remains to be seen if
observations indicate whether or not a potential with a minimum is favored.
With the availability of more precise data in future, the accurate 
reconstruction of quintessence potentials and the equation of state 
parameter may be possible. It is hoped that programmes such as the ones
initiated in Ref.\cite{saini} using red shift - luminosity distance 
correlations and in Ref.\cite{doran} using CMB data\cite{lee} will
enable the enforcement of tighter constraints on the signature and value
of the parameters $A$ and $\b$.

\section{Conclusions}

We have analysed the  dynamics arising from a scalar field rolling 
down the slope of a exponential potential in the framework of brane cosmology.
The brane world inflationary scenario is feasible with steep potentials as
distinct from the situation in standard cosmology. This is the case for
both exponential potentials\cite{copeland} and inverse power law
potentials\cite{huey}. During inflation
the desiradata of enough inflation and the COBE normalised amplitude
of density perturbations are used to fix the values of the brane tension
and the scale of the potential at this stage.The 
generic predictions of the models using exponential potentials is the
parameter independence of the spectral index $n$ and the tilt 
$A_S^2/A_T^2$\cite{copeland}. Subsequently, the requirements of suppression
of gravitino production and the emergence of radiation domination before
nucleosynthesis constrain the effective slope of the potential after inflation.
The tracker behaviour of the scalar field in an exponential potential ensures
the viability of dynamics in the matter dominated era.

The emergence of a second phase of accelerated expansion that we observe today,
necessitates the introduction of a second exponential term in the potential.
From the point of view of construction of a model of just quintessence, there
exists an ongoing debate as to whether\cite{cline2} or not\cite{barreiro,rubano}
a potential with a single exponential term is able to produce a viable
scenario. However, our analysis shows that a combination of two exponential
terms is essential for obtaining both inflation and present acceleration. The
motivation for considering such a potential is largely phenomenological.
Nevertheless, such kind of potentials do arise in the conformal Einstein
frame due to Brans-Dicke or other
types of nonminimal couplings of the scalar field. The inclusion of a higher
dimensional quantum effect or a cosmological constant together with a four
dimensional potential gives rise to effective potentials with two or more
exponential terms\cite{majumdar}. 

In order to obtain viable inflation and radiation and matter dominated eras we
ensure that one of the two terms in the potential dominates throughout these
epochs. The values of the parameters must be so chosen that the second term
starts playing a comparable role after the era of galaxy formation. In this
way we are able to obtain a workable scheme of quintessence. To conclude, we
obtain a scalar field dominated cosmology in the brane world scenario 
where one part of the potential
drives inflation, and the other part plays a crucial role in quintessence. In
order to achieve this with three parameters $(\a,\b,A)$ in the potential, our
results show that through the present status of analysis of 
observational data\cite{bernandis,perlmutter,lee} one parameter
($\a$) is constrained, another (either $A$ or $\b$) has to be tuned, and the
sign of the remaining one is dictated by the choice of the other two.

\end{multicols}


\begin{thebibliography}{99}

\bibitem{rubakov}
V. Rubakov and M. E. Shaposhnikov, Phys. Lett. B{\bf 159}, 22 (1985); J.
Polchinsky, Phys. Rev. Lett. {\bf 75}, 4724 (1995); P. Horava and E. Witten, 
Nucl. Phys. B{\bf 460}, 506 (1996); N. Arkani-Hamed, S. Dimopoulos and 
G. Dvali, Phys. Lett. B{\bf 429}, 263 (1998); I. Antoniadis, N. Arkani-Hamed, 
S. Dimopoulos and G. Dvali, Phys. Lett. B{\bf 436}, 257 (1998); A. Lukas, 
B. A. Ovrut and D. Waldram, Phys. Rev. D{\bf 60}, 086001 (1999).

\bibitem{randall}
L. Randall and R. Sundrum, Phys. Rev. Lett. {\bf 83}, 3370 (1999); {\it ibid}
{\bf 83}, 4690 (1999); 
N. Arkani-Hamed, S. Dimopoulos, G. Dvali and N. Kaloper, Phys.
Rev. Lett. {\bf 84}, 586 (2000); A. Chamblin and G. W. Gibbons, Phys. Rev. 
Lett. {\bf 84}, 1090 (2000).

\bibitem{cline}
J. M. Cline, C. Grojean and G. Servant, Phys. Rev. Lett. {\bf 83}, 4245 (1999);
T. Shiromizu, K. Maeda and M. Sasaki, Phys. Rev. D{\bf 62}, 024012 (2000); P.
Binetruy, C. Deffayet and D. Langlois, Nucl. Phys. B {\bf 565}, 269 (2000);
R. N. Mohapatra, A. Perez-Lorenzana and  C. A. de S. Pires, Phys. Rev. D 
{\bf 62}, 105030 (2000); R. N. Mohapatra, A. Perez-Lorenzana and  
C. A. de S. Pires, Int. J. Mod. Phys. {\bf A}16, 1431 (2001); R. Maartens, 
gr-qc/0101059.

\bibitem{bernandis}
P. de Bernandis {\it et al.}, Nature {\bf 404}, 955 (2000); A. Balbi 
{\it et al.}, Ap. J. {\bf 545}, L1 (2000); A. E. Lange
{\it et al.}, Phys. Rev. D{\bf 63}, 042001 (2001).

\bibitem{lidsey}
For recent reviews see, J. E. Lidsey, A. R. Liddle, E. W. Kolb, E. J. 
Copeland, T. Barreiro and M. Abney, Rev. Mod. Phys. {\bf 69}, 373 (1997); D. H.
Lyth and A. Riotto, Phys. Rep. {\bf 314}, 1 (1999); A. R. Liddle and D. H. 
Lyth, {\it Cosmological Inflation and Large Scale Structure} (Cambridge 
University Press, Cambridge, 2000).

\bibitem{maartens}
R. Maartens, D. Wands, B. A. Bassett and I. P. C. Heard, Phys. Rev D{\bf 62},
041301 (2000).

\bibitem{huey}
G. Huey and J. Lidsey, astro-ph/0104006.

\bibitem{copeland}
E. J. Copeland, A. R. Liddle and J. E. Lidsey, astro-ph/0006421.

\bibitem{ford}
L. H. Ford, Phys. Rev. D{\bf 35}, 2955 (1987); L. P. Grishchuk and Y. V.
Sidorov, Phys. Rev. D{\bf 42}, 3413 (1990); B. Spokoiny, Phys. Lett. B{\bf 315},
40 (1993); M. Joyce and T. Prokopec, Phys. Rev. D{\bf 57}, 6022 (1998); G.
Felder, L. Kofman and A. Linde, Phys. Rev. D{\bf 60}, 103505 (1999).

\bibitem{peebles}
P. J. E. Peebles and A. Vilenkin, Phys. Rev. D{\bf 59}, 063505 (1999); P.
Ferreira and M. Joyce, Phys. Rev. D{\bf 58}, 023503 (1998).

\bibitem{perlmutter}
S. Perlmutter {\it et al.}, Nature {\bf 391}, 51 (1998); A. G. Riess {\it et al.},
Astron. J {\bf 116}, 1009 (1998); P. Garnavich {\it et al.}, Ap. J. {\bf 493},
L53 (1998); B. P. Schmidt {\it et al.}, {\it ibid.} {\bf 507}, 46 (1998).

\bibitem{ratra}
B. Ratra and P. J. E. Peebles, Phys. Rev. D{\bf 37}, 3406 (1988); P. J. E.
Peebles and B. Ratra, Astron. J. {\bf 325}, L17 (1988); C. Wetterich, Nucl.
Phys. B{\bf 302}, 668 (1998); R. R. Caldwell, R. Dave and P. J. Steinhardt,
Phys. Rev. Lett. {\bf 80}, 1582 (1998); L. Wang, R. R. Caldwell, J. P.
Ostriker and P. J. Steinhardt, Ap. J. {\bf 530}, 17 (2000).

\bibitem{zlatev}
I. Zlatev, L. Wang and P. J. Steinhardt, Phys. Rev. Lett. {\bf 82}, 896 (1999);
P. J. Steinhardt, L. Wang and I. Zlatev, Phys. Rev. D{\bf 59}, 123504 (1999).

\bibitem{copeland2}
See for instance, E. J. Copeland, A. R. Liddle and D. Wands, Phys. Rev. 
D{\bf 57}, 4686 (1998); A. R. Liddle and R. J. Scherrer, {\it ibid.} {\bf 59},
023509 (1999); S. Perlmutter, M. S. Turner and M. White, Phys. Rev. Lett.
{\bf 83}, 670 (1999); A. Albert and C. Skordis, Phys. Rev. Lett. {\bf 84},
2076 (2000); V. Sahni and L. Wang, Phys. Rev. D{\bf 62}, 103517 (2000).

\bibitem{doran}
M. Doran, M. Lilley and C. Wetterich, astro-ph/0105457.

\bibitem{barreiro}
T. Barreiro, E. J. Copeland and N. J. Nunes, Phys. Rev. D{\bf 61}, 127301 (2000).

\bibitem{liddle}
A. R. Liddle, A. Mazumdar and F. E. Schunck, Phys. Rev. D{\bf 58}, 061301 
(1998); E. J. Copeland, A. Mazumdar and N. J. Nunes, Phys. Rev. D{\bf 60},
083506 (1999).

\bibitem{bunn}
E. F. Bunn, A. R. Liddle and M. White, Phys. Rev. D{\bf 54}, 5917R (1996);
E. F. Bunn and M. White, Ap. J. {\bf 480}, 6 (1997).

\bibitem{langlois}
D. Langlois, R. Maartens and D. Wands, Phys. Lett. B{\bf 489}, 259 (2000).

\bibitem{sarkar}
S. Sarkar, Rep. Prog. Phys. {\bf 59}, 1493 (1996).

\bibitem{allahverdi}
R. Allahverdi, A. Mazumdar and A. Perez-Lorenzana, hep-ph/0105125.

\bibitem{bean} 
R. Bean, S. H. Hansen and A. Melchiorri, astro-ph/0104162.

\bibitem{johri}
V. B. Johri, astro-ph/0007079.

\bibitem{bahcall}
N. A. Bahcall, J. P. Ostriker, S. Perlmutter and P. J. Steinhardt, Science,
{\bf 284}, 1481 (1999).

\bibitem{cline2}
J. M. Cline, hep-ph/0105251.

\bibitem{saini}
T. D. Saini, S. Raychaudhuri, V. Sahni and A. A. Starobinsky, Phys. Rev. Lett.
{\bf 85}, 1162 (2000).

\bibitem{lee}
A. T. Lee, astro-ph/0104459; C. B. Netterfield {\it et al.}, astro-ph/0104460.

\bibitem{rubano}
C. Rubano and P. Scudellaro, astro-ph/0103335.

\bibitem{majumdar}
A. S. Majumdar and S. K. Sethi, Phys. Rev. D{\bf 46}, 5315 (1992); A. S.
Majumdar, T. R. Seshadri and S. K. Sethi, Phys. Lett. B{\bf 312}, 67 (1993);
A. S. Majumdar, Phys. Rev. D{\bf 55}, 6092 (1997).


\end{thebibliography}
\end{document}